# The utility of infrasound in global monitoring of extraterrestrial impacts: A case study of the 23 July 2008 Tajikistan bolide

Elizabeth A. Silber[1,*]

[1]Geophysics, Sandia National Laboratories, Albuquerque, NM, 87123 (esilbe[at]sandia.gov)

*Corresponding author

## Abstract

Among various observational techniques used for detection of large bolides on a global scale is a low frequency sound known as infrasound. Infrasound, which is also one of the four sensing modalities used by the International Monitoring System (IMS), offers continuous global monitoring, and can be leveraged towards planetary defense. Infrasonic records can provide an additional dimension for event characterization and a distinct perspective that might not be available through any other observational method. This paper describes infrasonic detection and characterization of the bolide that disintegrated over Tajikistan on 23 July 2008. This event was detected by two infrasound stations at distances of 1530 and 2130 km. Propagation paths to one of the stations were not predicted by the model despite being clearly detected. The presence of the signal is attributed to the acoustic energy being trapped in a weak but leaky stratospheric AtmoSOFAR channel. The infrasound signal analysis indicates that the shock originated at the point of the main breakup at an altitude of 35 km. The primary mode of shock production of the signal detected at the two stations was a spherical blast resulting from the main gross fragmentation episode. The energy estimate, based on the signal period, is 0.17–0.51 kt of TNT equivalent, suggesting a mass of 6.6–23.5 tons. The corresponding object radius, assuming the chondritic origin, was 0.78–1.18 m.

## 1. Introduction

Observations of large bolide impacts on a global scale are of utmost importance for a number of reasons, including planetary defense, impact risk hazard assessment, event characterization, and improving global monitoring efforts (e.g., Chapman 2008; Christie & Campus 2010; Mainzer 2017; Silber et al. 2018; Trigo-Rodríguez 2022). Current monitoring technologies include ground instruments (e.g., dedicated camera systems, infrasound, radar) (e.g., Janches et al. 2006; Koschny et al. 2017; Silber & Brown 2019; Drolshagen et al. 2021), floating platforms (e.g., scientific payloads suspended on high-altitude balloons) (Young et al. 2018; Bowman 2021), and space-based assets (e.g., US government sensors, Geostationary Lightning Mapper (GLM)) (Nemtchinov et al. 1997; Jenniskens et al. 2018; Ott et al. 2021; Morris et al. 2022). These technologies can work in unison or independently, depending on circumstances, geographical coverage, energetics of the event, and data availability. Further to this, some of these technologies are more 'mature' (e.g., video camera detections) than others (e.g., McCrosky & Ceplecha 1968; Ceplecha 1977; Hawkes et al. 1984). Balloon infrasound, for example, while a promising method for Earth-based surveillance as well as future extraterrestrial exploration (Bowman 2021; Krishnamoorthy & Bowman 2023), is still considered an emerging technology. On the other hand, ground-based infrasound surveillance has been around off and on since the early 1960s (see Section 2) (Revelle 1997; Christie & Campus 2010). Lastly, while some instruments can provide unprecedented ground truth (GT) information (e.g., video) (Devillepoix et al. 2020), data collected by other types of instruments (e.g., infrasound and seismic) are best leveraged when GT collected by other means already exists (Silber & Brown 2019; Pilger et al. 2020). Nevertheless, it is remarkably difficult to obtain all parameters about the source, except in very exceptional circumstances when there are ample multi-modal observational data as well as physical samples available (e.g., Jenniskens et al. 2012; Brown et al. 2013).

Small near-Earth asteroids (NEAs) can pose a significant threat (Boslough et al. 2015; Harris & Chodas 2021); therefore, it essential to derive accurate parameters through detailed observations using any and all available sensing modalities. Integral to this endeavor are data fusion methodologies and conducting model cross-validation using different techniques in order to verify the reliability and robustness of derived parameters.

A significant challenge in characterizing bolides comes from their diversity (e.g., size, composition, porosity, velocity, etc.) and impact randomness (i.e., could impact anywhere, anytime on the globe). In principle, no two bolide events are alike, nor can we meaningfully anticipate where they will impact. Only recently have there been cases of an imminent impact announced only a few hours ahead of the event (Clark et al. 2023) but that does not allow sufficient time for well-planned, comprehensive and dedicated observations (Silber et al. 2023a). Although models have been developed to derive bolide parameters even when limited observational evidence is present (e.g., see Sansom et al. 2019 and references within), it is not always possible nor feasible to obtain reliable and/or comprehensive GT information for all events of interest or those events that might not be detected by preferred sensing modalities. Despite the continuous advancements in theoretical models, there are still uncertainties stemming from the fact that some initial assumptions must be made in order to apply models in the first place.

In recent years, especially since the spectacular Chelyabinsk event (Brown, Assink et al. 2013; Popova et al. 2013), there has been an increased interest in leveraging infrasound towards bolide detection, geolocation, and characterization. However, despite this, the number of case studies presented and available in literature remains underwhelmingly low compared to bolide characterization through other means (e.g., optical).

This paper presents infrasound detections and signal analysis of the Tajikistan bolide, which impacted just after the sunset local time on 23 July 2008. The remainder of the paper is organized as follows: Section 2 presents background on bolide infrasound and other modes of observations that provide GT information. In Section 3, the observations and GT for the Tajikistan bolide are presented. Infrasound analysis is outlined in Section 4, and the results and discussion are presented in Section 5. The conclusions are given in Section 6.

## 2. Bolide infrasound

Entry velocities ($v$) of extraterrestrial objects range from 11.2 – 72.8 km/s (Ceplecha et al. 1998), which translate to Mach numbers 35 – 270 (Mach number ($M$) is a ratio of the object's speed to the local speed of sound ($c_s(z)$)) (Ben-Dor et al. 2000; Silber, Boslough et al. 2018). Velocities greater than ~73 km/s are associated with objects originating beyond the confines of our Solar

System, such as the recent interstellar 'visitor' 'Oumuamua (Meech et al. 2017). However, in some exceptional cases, excess velocity might be explained solely by gravitational forces, even if the object belongs to the Solar System, as proposed by Peña-Asensio et al. (2024). The authors presented an example of a meteoroid with a measured velocity of 73.7 ± 0.6 km/s, and originating from within the Solar System.

Any object moving at a speed greater than the local speed of sound will produce a 'sonic boom' wave, with energy confined to a Mach cone with a half angle corresponding to $sin^{-1}\left(\frac{v(z)}{c_s(z)}\right)$ and propagating outward at approximately the speed of sound. Here, *v*(*z*) is the object speed at some altitude, and $c_s(z)$ is the local speed of sound at that altitude. The prerequisite is that the object must be in the continuum or slip flow regime, where the local Knudsen number (*Kn*) is ≤ 0.001 and 0.001 < *Kn* ≤ 0.1, respectively (Anderson 2000). In bolides, since they travel at high speeds, the Mach cone angle is so small that a moving source can be approximated as a cylindrical line, where a blast wave propagates perpendicular to the trajectory (Plooster 1970; Tsikulin 1970; ReVelle 1976; Ben-Dor, Igra et al. 2000). However, the bolide entry geometry and shock production mechanisms can complicate things; if the entry angle is steep, refractive effects can be significant, as shown by ReVelle (1976), and if there is fragmentation, a cylindrical line source no longer applies.

The region where highly non-linear effects dominate, also known as the characteristic or relaxation radius ($R_0$) of the cylindrical blast wave is proportional to the square root of the energy deposited per unit length ($E_L$):

$$R_0 = \left[\frac{E_l(z)}{p_s(z)}\right]^{1/2} \approx M d_i \qquad (1),$$

where $p_s(z)$ is the ambient pressure at source altitude (z), and $d_i$ is the impactor diameter. The assumption here is that there is no fragmentation. In the case of gross fragmentation, a spherical shock will form, with the characteristic radius proportional to the cube root of the energy ($E_S$) released by the explosion:

$$R_0 = \left[\frac{E_S(z)}{\frac{4\pi}{3} p_s(z)}\right]^{1/3} \qquad (2).$$

Beyond approximately $10R_0$, the highly non-linear shock will transition to a weak shock, with the fundamental frequency ($f_0$) defined as:

$$f_0 = \frac{1}{\tau_0} = \frac{c_s(z)}{2.81 R_0} \quad (3),$$

where $\tau_0$ is the fundamental period of the wave (ReVelle 1976). Attenuation and dispersion act to reduce the amplitude and stretch the period (Dumond et al. 1946), until the wave eventually transitions into a linear regime. For a comprehensive overview of bolide infrasound, the reader is directed to a book chapter by Silber and Brown (2019).

A fraction of energy released by a bolide will inevitably go into acoustic energy of the remnant shock. This acoustic energy is typically in the infrasonic range, with frequencies below 20 Hz. Other than bolides, there are many other sources of infrasound, including explosions (e.g., Arrowsmith et al. 2008; Ceranna et al. 2009), volcanoes (e.g., Matoza et al. 2019), re-entry (e.g., ReVelle et al. 2005; Silber, Bowman et al. 2023a), lightning (e.g., Assink et al. 2008; Farges 2009), and rockets (e.g., Pilger et al. 2021; Albert et al. 2023). Some of these might have similar frequency content and similar signatures, making the source identification using infrasound records alone challenging.

Infrasound generated by bolides, especially if they are sufficiently large and fast, and penetrate deep into the atmosphere, can be detected at long distances (Silber et al. 2011; Brown, Assink et al. 2013). Depending on several factors, including atmospheric propagation effects, an acoustic wave signature should contain some information about the source (e.g., energy), as well as provide nominal diagnostic for the mode of shock production (a cylindrical blast wave during a hypersonic passage and/or a quasi-spherical or spherical explosion through fragmentation and/or terminal airburst) (Revelle 1997; Ceplecha, Borovička et al. 1998; Silber, Le Pichon et al. 2011).

The earliest documented infrasound generated by an extraterrestrial impact was that from the 1908 Tunguska airburst (Whipple 1930). Because acoustic waves at inaudible frequencies can propagate over large distances with little attenuation, carrying information about the source, infrasound has been used for global monitoring. From the early 1950s and up until the mid-1970s, infrasound played an important role in such endeavors, providing robust and low-cost global surveillance of explosive phenomena (Revelle 1997). Given the passive nature of infrasound monitoring, many other types of events were detected, including ten large bolides (Revelle 1997; Silber et al. 2009). The subsequent infrasound analyses some decades later were suggestive of an elevated impact risk (Silber, ReVelle et al. 2009), which was corroborated through other means shortly after the Chelyabinsk superbolide event (Brown, Assink et al. 2013). As infrasound surveillance dwindled into obscurity in the mid-1970s, so did the associated research. After the Comprehensive Nuclear-Test-Ban Treaty (CTBT) was opened for signature in the 1990s, infrasound once again entered the scientific scene, and it now serves as one of the four monitoring technologies of the International Monitoring System (IMS). There are 53 fully installed and certified infrasound stations (out of planned 60), providing global coverage for detections of anthropogenic and natural phenomena, including bolides (Figure 1).

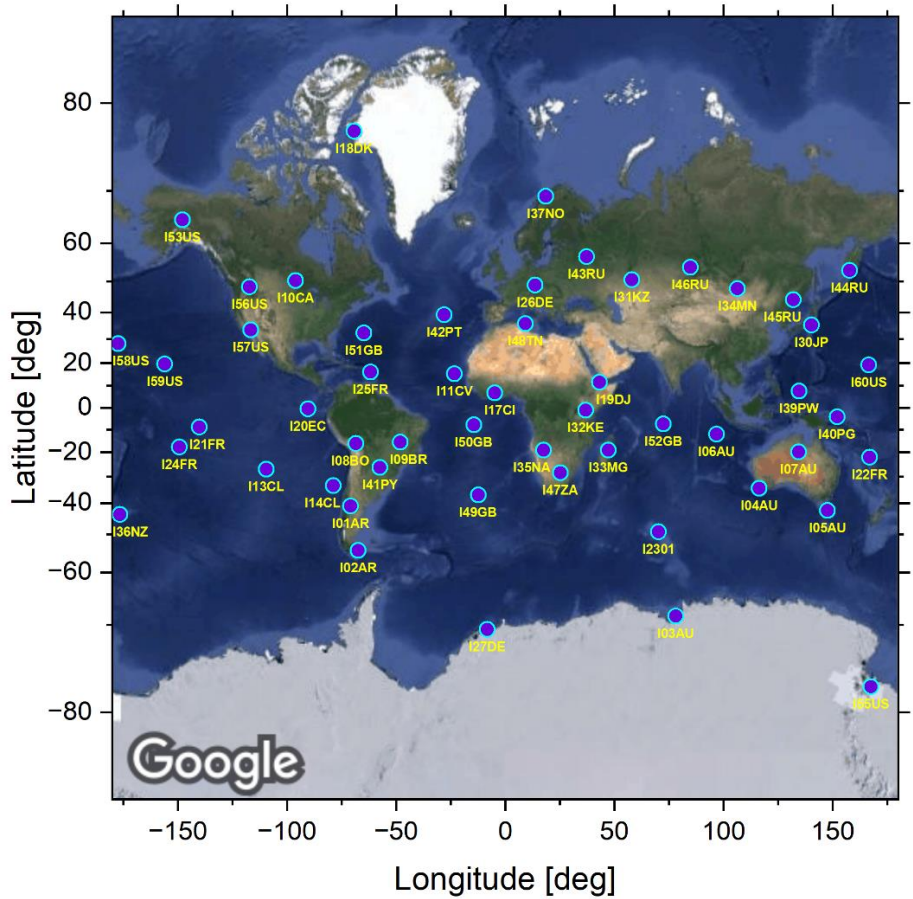

**Figure 1:** Global map showing the fully operational and certified infrasound stations of the IMS. The map is Mercator projected.

In order to identify and characterize a bolide event via infrasound, it is crucial to have some GT information. Ideally, this would include timing and location, as well as the altitude and velocity at the very least. Ground-based observations (e.g., casual witnesses, video cameras) are confined to narrow geographical regions, making it impossible to achieve global coverage. Space assets can provide near continuous global monitoring and detections of bolides. However, a system dedicated specifically to bolide monitoring and detections does not exist. The GLM instrument onboard the Geostationary Operational Environmental Satellites (GOES-16 and GOES-17), designed to detect lightning activity from space, has been shown to be capable of detecting bright fireballs in a narrow passband (Jenniskens, Albers et al. 2018). The geographical coverage of GLM is confined to the region of the Americas and a limited band of the Atlantic and Pacific oceans.

Another space-based asset includes the US government (USG) sensors, which can detect bolides globally (Tagliaferri et al. 1994). While not all bolides that impact Earth at any time and location are detected and reported, the list published by the Jet Propulsion Laboratory's (JPL) Center for Near Earth Object Studies (CNEOS) contains a large number of events with GT information which can be utilized towards infrasound characterization of bolides (https://cneos.jpl.nasa.gov/fireballs/). Since April 1988 and as of early February 2024, there have been 853 fireball detections by the USG sensors. All events have a date and time, the total radiated energy (in joules), and the calculated total impact energy (in kt of TNT equivalent; 1 kt = $4.184 \cdot 10^{12}$ J). Most events (782, or 91.7%) also have a precise geographical location, with latitudes spanning from the northern to the southern polar regions (Figure 2).

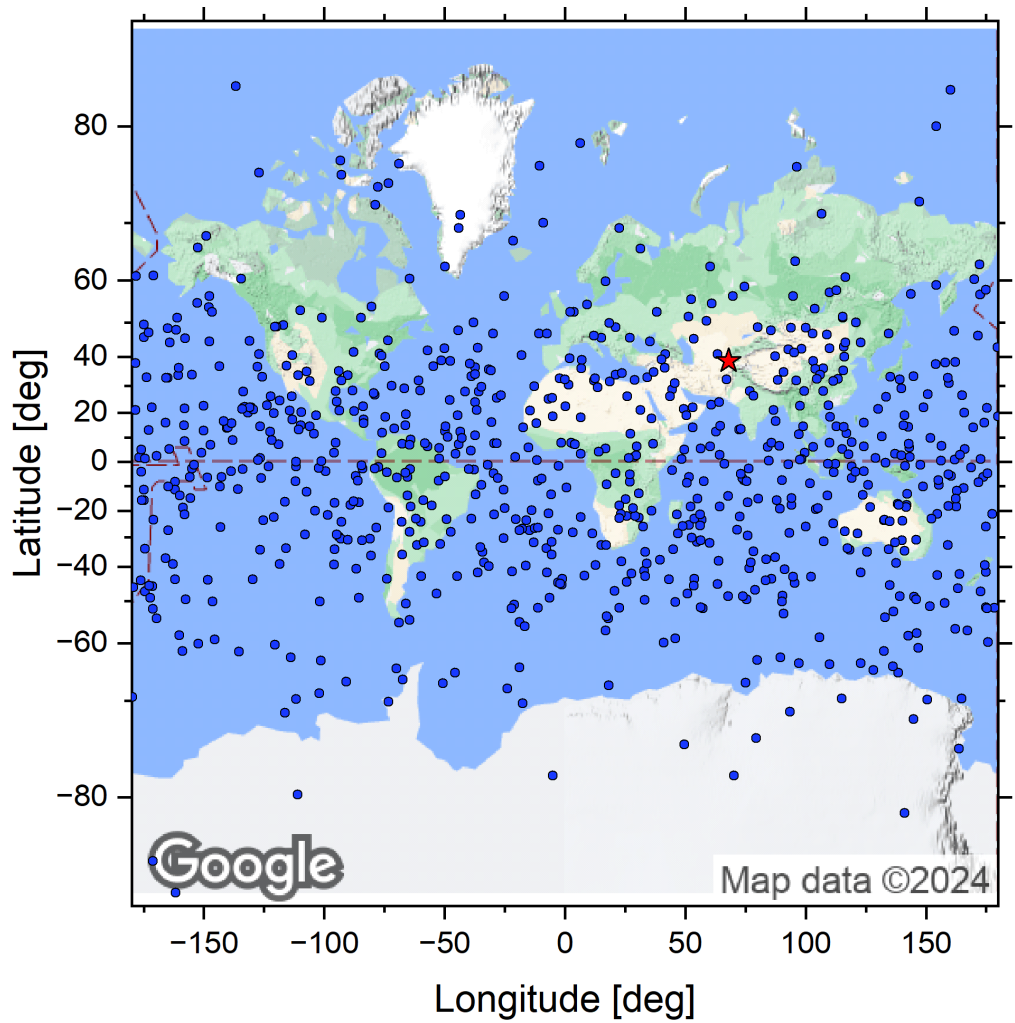

**Figure 2:** Map showing all bolide detections as listed in the CNEOS database as of February 2024. There are 782 events with a reported geographical location. The Tajikistan bolide is shown with the red star. The map is Mercator projected.

The peak brightness altitude is known for 513 events and the pre-atmospheric velocity with *x*-, *y*-, and *z*-components for 304 fireballs. Only three fireballs have a velocity vector but not the altitude of peak brightness. Previous studies reported inaccuracies in the reported velocities for some events (e.g., Borovička et al. 2017; Devillepoix et al. 2019). In terms of energy (*E*), most fireballs (865) have their yields in the sub-kt range (<1 kt). Only ten events exceed 10 kt, with the Chelyabinsk superbolide dominating the total population with its energy of 440 kt (Brown, Assink et al. 2013). Below 10 kt, there are 15 events with energy $5\text{kt} \leq E < 10\text{kt}$, 29 events with energy $2\text{kt} \leq E < 5\text{kt}$, and 50 events with energy $1\text{kt} \leq E < 2\text{kt}$. As of last year, the light curve data became available as well, including visible light intensity (in watts per steradian) and total radiated energy (in joules) as a function of time. While altitudes are not available, the light curve data are very useful in establishing whether there was any fragmentation, how many such episodes, and if these occurred throughout the flight or are skewed towards beginning or the end of the visible light trail. Figure 3 shows how the light curve and the corresponding infrasound signal might look like in the two contrasting scenarios – without (left panels) and with fragmentation (right panels). Although this is an idealized depiction, it illustrates the interconnectedness of various features in light curves and infrasound signals, and how these might aid in interpretation and source characterization. However, an important note about infrasound must be made here. As an acoustic wave propagates over long distances, dispersion, scattering, turbulence, prevailing winds, and other atmospheric effects can alter the signal so much that it no longer contains obvious source-identifying features, making analysis a challenging task.

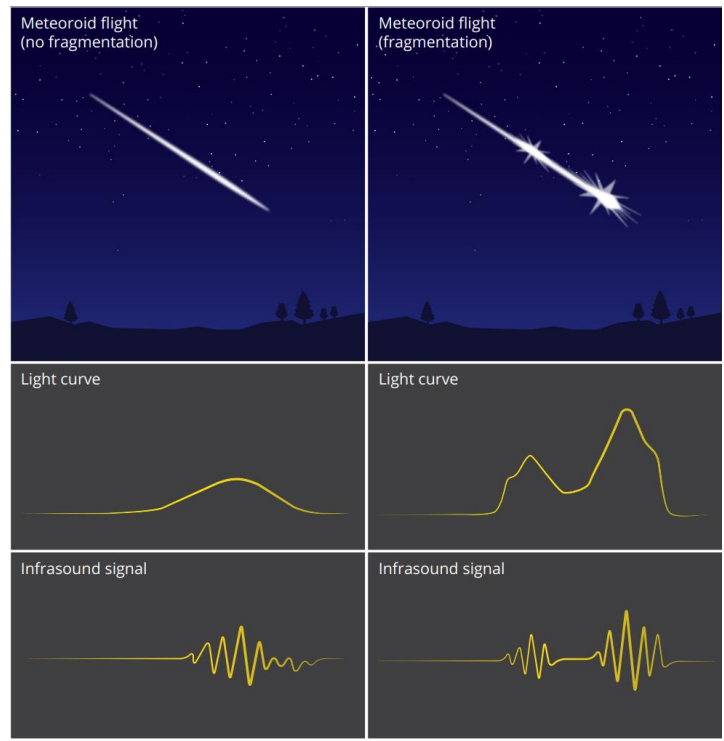

**Figure 3:** Diagram showing the idealized light curve and infrasound signatures resulting from the scenarios without fragmentation (left panels) and with fragmentation (right panels).

## 3. The Tajikistan bolide

On 23 July 2008 at 14:45:25 UTC (19:45:25 local time) a very bright object entered the atmosphere, producing a series of audible sounds, as reported by casual witnesses. Thunder-like sounds, including a sonic boom, most likely generated during the bolide fragmentation, were heard as far as 100 km from the burst site (Konovalova et al. 2013). Also remarkable was that the bright flash from the bolide was also seen at that distance, and even noticed by witnesses who were indoors. The authors reported that the most prominent sonic booms were heard approximately 30 km from the epicenter. Many casual witnesses also photographed a persistent smoke trail, which lingered for some 20 minutes. The bolide entered at a steep angle of 72.3° degrees, and at a relatively slow speed of 14.5 km/s. Table 1 outlines all known bolide parameters; these come from the CNEOS database, Konovalova, Madiedo et al. (2013), and Peña-Asensio et al. (2022). The bolide was detected by five seismic stations as well (Konovalova, Madiedo et al. 2013).

The light curve, shown in Figure 4, was digitized from the data published in the CNEOS database. Using the peak brightness of $\sim 3.6 \times 10^{10}$ W/str from the light curve (Figure 4), and applying the conversion factor for a 6000 K black body model published by (Ceplecha, Borovička et al. 1998), the peak absolute visual magnitude was -20.3. This places the Tajikistan event into a superbolide category (also see Konovalova, Madiedo et al. (2013)). The light curve shows three peaks (also known as flares), one large and two smaller ones, corresponding to a sudden increase in brightness. There are a couple of reasons for a flare to occur – one is a fragmentation episode, and the other is a change in physical conditions, resulting in increased ionization, excitation and evaporation (Ceplecha, Borovička et al. 1998). The flares seen in Tajikistan bolide light curve are interpreted as progressive fragmentation episodes (Konovalova, Madiedo et al. 2013 and references therein). The first flare was the most prominent, indicative of gross fragmentation.

Konovalova, Madiedo et al. (2013) reconstructed the trajectory based on photographic records collected by casual witnesses, and by performing nighttime stellar calibration. This is one of the more challenging approaches to obtain the trajectory. Integrating puzzle pieces coming from infrasound, seismic, and USG sensor data, it was possible to establish some constraints that enabled reasonably robust results. This event illustrates the importance of high-fidelity observations and data fusion in accurate characterization of bolide events.

**Table 1:** Superbolide GT data. The parameters come from the following sources: (a) the CNEOS database, (b) Konovalova, Madiedo et al. (2013), and (c) Peña-Asensio, Trigo-Rodríguez et al. (2022). Note that the entry velocity comes from two sources, as listed in the table.

| Parameter | Value |
| --- | --- |
| Event time [UTC] (a) | 14:45:25 |
| Entry velocity [km/s] (a) | 14.5 |
| Entry velocity [km/s] (b) | 14.3 ± 0.5 |
| Event latitude and longitude [°] (a) | 38.6 N, 68.0 E |
| Entry angle [°] (c) | 72.3 |
| Azimuth angle [°] (c) | 278.8 |
| Peak brightness altitude [km] (a) | 31.5 |
| Velocity components, $v_x$, $v_y$, $v_z$ [km/s] (a) | -7.7, -8.2, -9.1 |
| Total radiated energy [J] (a) | $12.1 \times 10^{10}$ |
| Total impact energy [kt] (a) | 0.36 |
| Absolute visual magnitude (b) | -20.3 |
| Initial observed altitude [km] (b) | 38.2 ± 0.5 |
| Terminal altitude [km] (b) | 19.6 ± 0.5 |
| Velocity at terminal altitude [km/s] (b) | 8.7 ± 0.5 |
| Primary flare (main breakup) altitude [km] (b) | 35 |
| Second flare altitude [km] (b) | 28 |
| Third flare altitude [km] (b) | 25 |
| Estimated radius [m] (b) | 1.15 ± 0.05 |
| Estimated mass [kg] (c) | $1.41 \times 10^{4}$ |
| Dynamic strength [MPa] (c) | 1.325 |

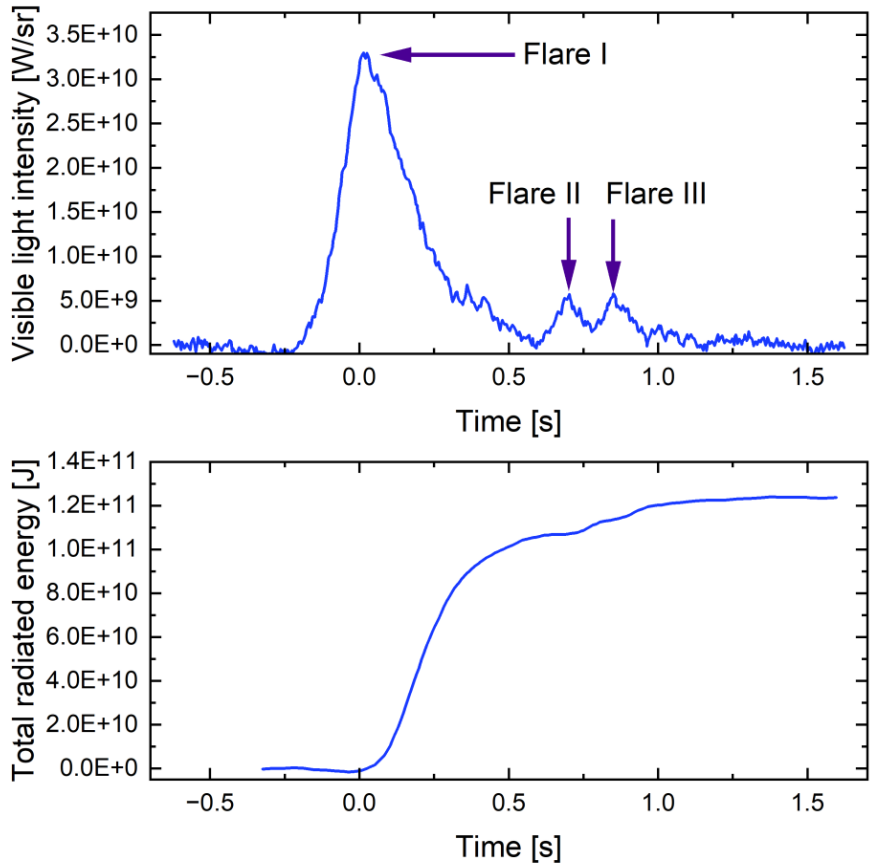

**Figure 4:** Bolide light curve (top panel) and total radiated energy (bottom panel) as a function of time. The plot data were digitized from the published light curve file posted on the CNEOS database (webpage).

As per the analysis of Konovalova, Madiedo et al. (2013), the luminous path extended from an altitude of 38.2 km down to 19.6 km, with the main breakup occurring at an altitude of 35.0 km. They also indicated that the second flare likely occurred at 28 km. As per the CNEOS database, however, the peak brightness altitude was 31.5 km. The object was determined to be of an asteroidal origin, likely about 2 meters in diameter and with the pre-atmospheric mass of 20–25 tons. Even though the bolide penetrated deep into the atmosphere, no meteorites were ever found (Konovalova, Madiedo et al. 2013). Peña-Asensio, Trigo-Rodríguez et al. (2022) performed the orbit reconstruction for 255 events listed in the CNEOS database, including the Tajikistan bolide. Table 1 includes entry and azimuth angles, as well as the dynamic strength of the Tajikistan impactor. The latter quantity agrees well with that determined by Konovalova, Madiedo et al. (2013). Gi et al. (2018) used the analytic Triggered Progressive Fragmentation Model (TPFM) (ReVelle 2005) to model simple fragmentation using a handful of bolide events (including the Tajikistan superbolide) in order to estimate how often such events might cause window breakage on the ground. They were able to match the main flare altitude reported by Konovalova, Madiedo et al. (2013). The mass and radius Gi, Brown et al. (2018) obtained are in line with that reported by Peña-Asensio, Trigo-Rodríguez et al. (2022).

An initial report noting infrasound detections was presented at a conference by Popova et al. (2011). The signal measurements were also obtained by Ens et al. (2012) as part of a larger database study and subsequently reported by Gi and Brown (2017). This paper presents in-depth signal analysis, source characterization, and propagation modeling.

## 4. Methods

A search for signals at IMS infrasound stations up to 5000 km was carried out. Two software packages were used to search for detections, the Progressive Multi-Channel Correlation Method (PMCC) algorithm (Cansi 1995; Brachet et al. 2010), and MatSeis (Harris & Young 1997; Young et al. 2002). The PMCC technique is sensitive to coherent signals with a very low signal-to-noise ratio (SNR) and has been successfully employed in examining infrasound from other large bolides (e.g., Silber, Le Pichon et al. 2011). The two closest stations, I31KZ and I46RU, at distances of 1530 km and 2130 km, respectively, detected the signals consistent with the superbolide event at 38.6°N, 68.0°E. Figure 5 shows the locations of the infrasound stations, as well as the bolide.

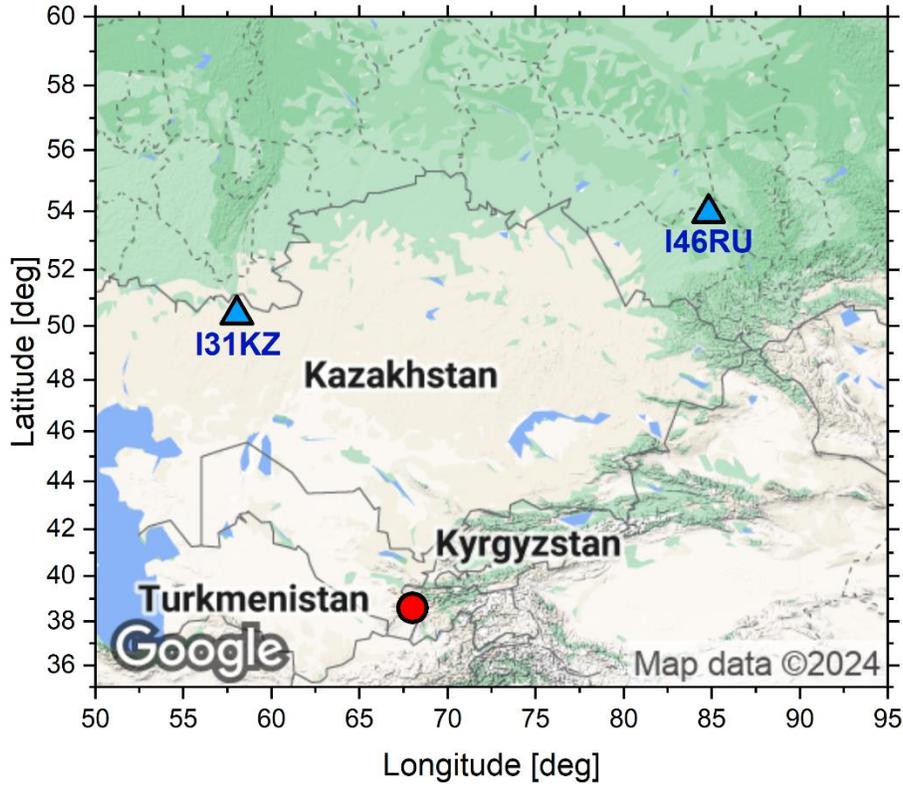

**Figure 5:** Map showing the locations of the bolide event and the two infrasound stations.

Once beamforming was performed and the airwave arrival parameters determined, the timeseries were analyzed using a standalone procedure outlined in Ens et al. (2012). The maximum peak-to-peak amplitude was determined by bandpassing the stacked, raw waveform using a second-order Butterworth filter and then applying the Hilbert Transform (Dziewonski and Hales, 1972) to obtain the peak of the envelope. The period at maximum amplitude was computed by measuring the zero crossings of the stacked waveform at each station (Revelle 1997; Ens, Brown et al. 2012). The filter bandwidth was determined by overlaying a segment containing the signal and a segment containing the noise (the average between the prior and post noise) in the frequency domain and selecting the range which lies above the noise (for further details, see Ens, Brown et al. (2012)).

As it can be seen from Figure 5, the acoustic wave propagation direction is northwest to I31KZ and northeast to I46RU. During summers, prevailing winds in the northern hemisphere are easterly (Webb 1966); therefore, it is expected that there would be a notable effect. To account for possible

Doppler shift due to prevailing winds, the wind speed vector along the great circle to each station was tabulated and used to Doppler-correct the signal periods (Ens, Brown et al. 2012). The atmospheric specifications were obtained from the United Kingdom Meteorological Office (UKMO) assimilated dataset (Swinbank & O'Neill 1994) up to ~65 km. The atmospheric specifications for altitudes beyond that came from the HMW95 (Horizontal Wind Model; (Hedin et al. 1996)) and the NRL-MSIS00 (Naval Research Laboratories – Mass Spectrometer and Incoherent Scatter Radar (Picone 2002)). Another source of atmospheric specifications is the Ground-2-Space (G2S) model (Drob et al. 2003), hosted by the National Center for Physical Acoustics (NCPA), but data for a desired time frame (14:00 – 16:00 UTC) on 23 July 2008 were not available.

Propagation modeling using InfraGA, the open source raytracing package (Blom 2014; Blom & Waxler 2017), was carried out with an aim to evaluate propagation paths and to estimate the likely source altitude. InfraGA can be downloaded from Github: https://github.com/LANL-Seismoacoustics/infraGA. InfraGA can perform computations in the Cartesian and spherical coordinates using a realistic atmosphere, with the shooting ray and eigenray search modes (Blom & Waxler 2017). Raytracing was done in 1 km increments at altitudes from 20 – 38 km.

There are several relations aimed at estimating yield. An extended background and discussion about these relations can be found in Silber and Brown (2019) and Ens, Brown et al. (2012). The most robust are the semi-empirical period-yield relations, originally derived for estimating nuclear explosion yield by the Air Force Technical Applications Center (AFTAC). These were later adapted for bolides by Revelle (1997) as follows:

$$\log(E) = 3.34 \log(\tau) - 2.28, \quad E \leq 200 \text{ kt} \quad (4a)$$

and

$$\log(E) = 4.14 \log(\tau) - 3.31, \quad E > 80 \text{ kt} \quad (4b).$$

Here, energy is in units of kt of TNT equivalent, and $\tau$ is the signal period measured at maximum amplitude. Ens, Brown et al. (2012) used 71 bolide events detected by IMS infrasound stations to refine the period-yield relation and arrived to the relations very similar to that of AFAC:

$$\log(E^*) = 3.75 \log(\tau) + 0.50 \quad (5a),$$

and

$$\log(E^*) = 3.28 \log(\tau_{avg}) + 0.71 \quad (5b).$$

Note that the energy ($E^*$) in these relations is in tons of TNT equivalent (Ens, Brown et al. 2012). The difference between Eq. (5a) and Eq. (5b) is the signal period used for calculations. The former uses the individual period as measured at each station for a given event, while the latter uses the averaged period across all stations for a given event. Gi and Brown (2017) extended the work of Ens, Brown et al. (2012), and analyzed 78 bolides in the context of CNEOS events, where various parameters, including altitude, velocity, and energy are available. They arrived at the following (in units of kt):

$$\log(E) = 3.68 \log(\tau) - 1.99 \quad (6a),$$

and

$$\log(E) = 3.84 \log(\tau_{avg}) - 2.21 \quad (6b).$$

## 5. Results and Discussion

Table 2 summarizes the measured signal properties for both stations. The signal arrived at I31KZ (located at 50°.41N, 58°.03E, 1530 km from the source), at 16:04:28 UTC, lasting about eight minutes. Both PMCC and MatSeis yielded the same results. An example of the output from MatSeis for I31KZ is shown in Figure 6. The onset of the signal at I46RU (located at 53°.45N, 84°.82E, 2130 km from the source) was at 16:42:35 UTC. The reverberations lasted around 6 minutes. While the dominant frequency is nearly the same at the two stations, 0.38 Hz at I31KZ and 0.33 at I46RU, the overall frequency content is quite different. The acoustic signals arriving at I46RU had a much lower frequency content (0.25–0.8 Hz) than those arriving at I31KZ (0.14–5.0 Hz). It is expected that higher frequences would be 'scrubbed out' out at larger distances (Southerland & Bass 2004; Norris et al. 2010), and I46RU is 600 km further away from the event than I31KZ. Furthermore, the atmosphere can have a profound effect on acoustic wave propagation (Southerland & Bass 2004; de Groot-Hedlin 2005; de Groot-Hedlin et al. 2010). Downwind propagation can amplify the signal by augmenting the amplitude and stretching the signal period. Conversely, upwind propagation will have the opposite effect. The locations of stations are such that there should be strong easterly crosswind; the cumulative effect along the propagation path indicates downwind conditions towards I31KZ (22.6 m/s) and upwind conditions (-16.7 m/s) towards I46RU. This could plausibly explain why only very low frequency content is seen at I46RU. Previous studies noted observing episodic upwind arrivals when such are not expected (Nippress & Green 2017). In particular, upwind arrivals exhibited low amplitude and low frequency content, consistent with the signals observed at I46RU. Other possible contribution might be the station noise (Bowman et al. 2005; Pilger et al. 2015), so the signal-to-noise (SNR) was evaluated as well. Both stations show similar signal SNR, and similar background noise levels.

**Table 2:** Signal properties.

| Parameter | I31KZ | I46RU |
|---|---|---|
| Station latitude [°] | 50.41 | 53.96 |
| Station longitude [°] | 58.03 | 84.82 |
| Signal arrival time [UTC] | 16:04:28 | 16:42:35 |
| Signal travel time [s] | 4647 | 6875 |
| Duration [s] | 479 | 352 |
| Distance [km] | 1529.5 | 2129.5 |
| Observed back azimuth [°] | 144.8 | 228.1 |
| Theoretical back azimuth [°] | 145.2 | 223.6 |
| Celerity [m/s] | 329 | 310 |
| Amplitude, P2P [Pa] | 0.153 ± 0.038 | 0.054 ± 0.018 |
| Amplitude, max [Pa] | 0.109 ± 0.019 | 0.031 ± 0.009 |
| Period, zero crossings [s] | 3.11 ± 0.12 | 2.71 ± 0.01 |
| Doppler corrected period [s] | 2.91 | 2.85 |
| Period, PSD [s] | 2.63 | 3.0 |
| Wind vector [m/s] | 22.6 | -16.7 |
| Trace velocity [m/s] | 357 | 339 |
| Peak frequency [Hz] | 0.38 | 0.33 |
| Frequency cutoffs [Hz] | 0.14 - 5.0 | 0.25 - 0.8 |

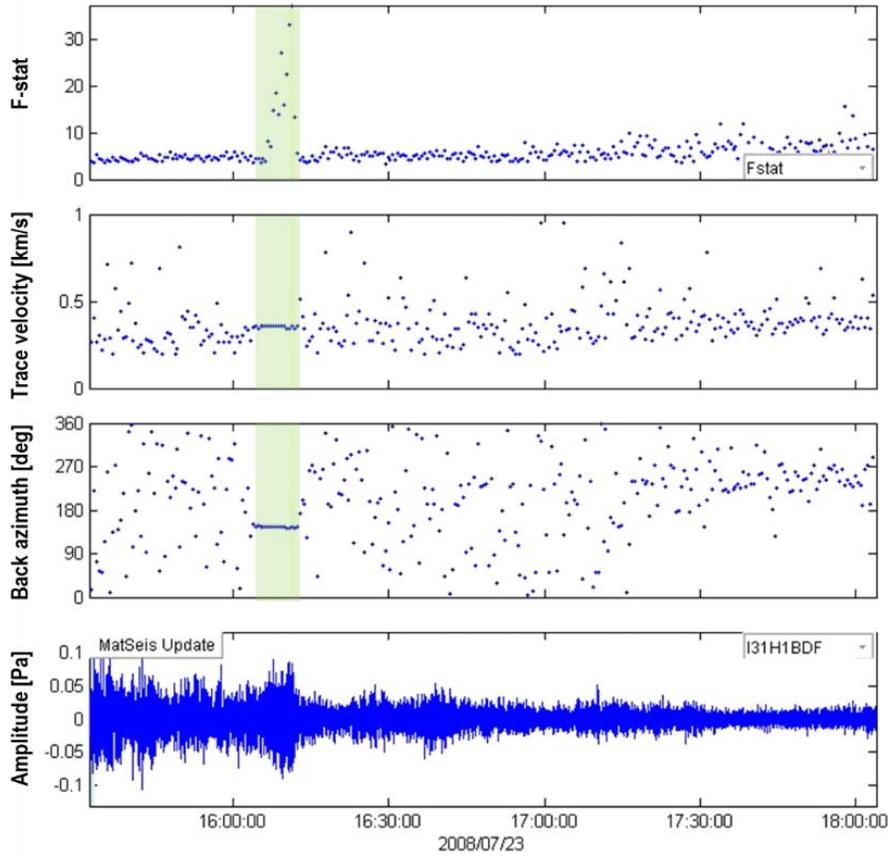

**Figure 6:** MatSeis output showing the signal arrival at I31KZ. The panels, from top to bottom are Fstat, trace velocity (in km/s), back azimuth, and the timeseries at Channel 1.

It is not unusual to observe a significant variation in measured signal periods across different stations (Silber, ReVelle et al. 2009; Silber, Le Pichon et al. 2011). Such pattern has been attributed to several possible factors, one of them being that infrasound stations are 'sampling' signals emanating from different parts of the trail, which might also be associated with different modes of shock production as the object interacts with the atmosphere during its descent. For example, a cylindrical line source generated by a given object will inevitably have a smaller blast radius (i.e., shorter dominant period, higher dominant frequency) compared to a quasi-spherical or spherical source resulting from gross fragmentation of that same object. Even if gross fragmentation never takes place, the region of maximum energy deposition will still have a larger blast radius (Zinn et al. 2004; Silber & Brown 2019). The peak brightness is associated with the highest energy deposition, and therefore the largest blast radius, which should be reflected in the signal measurements. The signal period (zero crossings method), adjusted for Doppler shift, was 2.91 s at I31KZ and 2.85 s at I46RU, which is consistent with a singular source, i.e., coming from approximately the same altitude. Based on the light curve data, as well as optical observations reported by Konovalova, Madiedo et al. (2013), it is highly likely that the source of both signals was the primary gross fragmentation episode. The peak brightness altitude listed in the CNEOS database is 31.5 km, while Konovalova, Madiedo et al. (2013) obtained the main flare (gross fragmentation) to have occurred at an altitude of 35 km. This apparent discrepancy might come from optical calibration uncertainties (Konovalova, Madiedo et al. 2013) or from inaccuracies in the reported CNEOS peak brightness altitudes. Using several representative large bolides as case studies, Brown et al. (2016) found that the peak brightness height listed on the CNEOS webpage is generally accurate to within ~3 km.

While this height difference as reported by Konovalova, Madiedo et al. (2013) and the CNEOS database is small relative to the propagation distance of the acoustic wave, the source altitude is relevant in the context of formation of acoustic ducts (Nippress & Green 2017; Albert, Bowman et al. 2023). Raytracing results for both stations, with the source originating at 35 km altitude are shown in Figure 7. The runs were done with 0.5° launch angle increments but to better visualize propagation path, rays in 5-degree increments are plotted.

There is a well-developed stratospheric duct between the source and I31KZ. The fastest predicted rays originate at altitudes between 32–38 km, with travel times underestimated by up to 4%. These altitudes reasonably correlate to the altitude of the main fragmentation event. It is plausible that the winds stronger than predicted by the model were present on that day. Moreover, while atmospheric specifications given by various models are generally robust, they do not include fine-scale perturbations or other local effects that might occur on scales shorter than the wave propagation (Chunchuzov et al. 2010; Green et al. 2011; Le Pichon 2015). It is likely that the atmospheric specifications do not fully capture inhomogeneities and fine scale variations, observable at scales of minutes and hours (Kulichkov et al. 2002; Chunchuzov, Kulichkov et al. 2010; Green, Vergoz et al. 2011; Averbuch et al. 2022). Such effects can lead to multipathing, and 'leakage' of energy from elevated ducts.

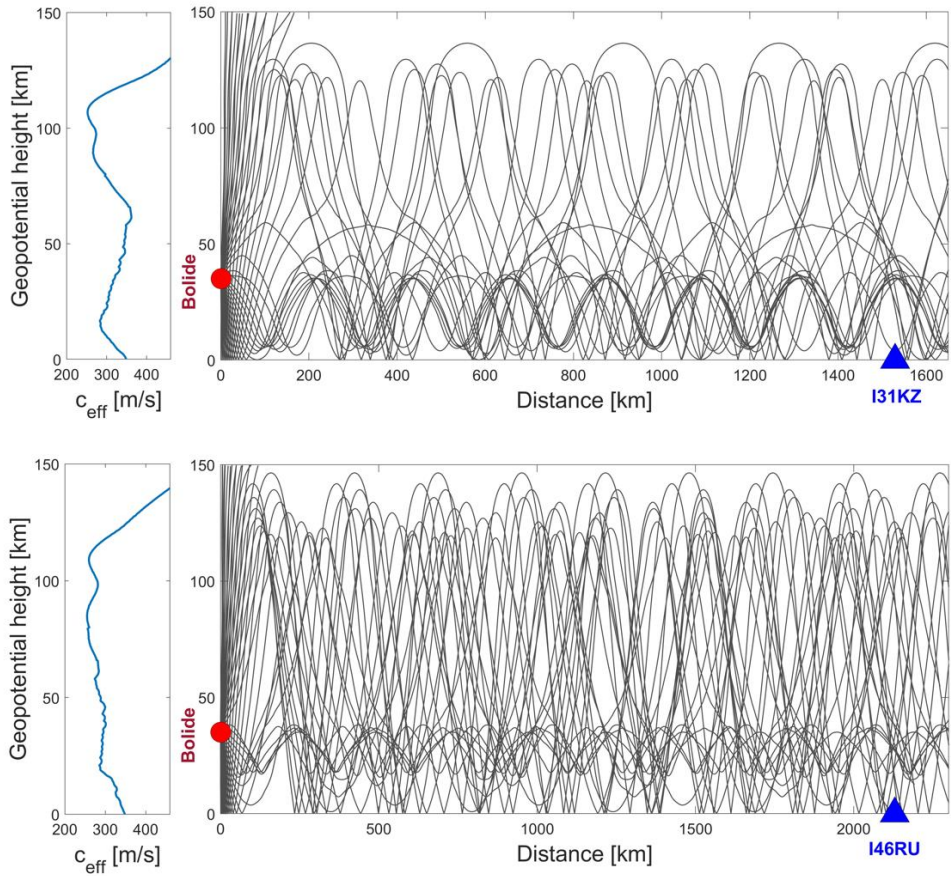

**Figure 7:** Raytracing results for I31KZ (top panel) and I46RU (bottom panel). The source is located an altitude of 35 km. The distance axes are scaled according to the location of the receiver. The effective sound speed in the direction of propagation is plotted on the left.

In the case of I46RU, the stratospheric duct is narrow and weak, and no stratospheric arrivals from any source altitude are predicted at the station, which contradicts the observational evidence. Only thermospheric arrivals are predicted but such were not observed. At this distance, thermospheric arrivals would be heavily attenuated and unlikely to be detected. Counter-wind arrivals are also difficult to model in general (de Groot-Hedlin, Hedlin et al. 2010). However, acoustic energy generated at altitude appears to be trapped in an elevated duct, consistent with a so-called AtmoSOFAR channel, which was modelled in the early 2000s (Drob, Picone et al. 2003) but only recently observationally confirmed (Albert, Bowman et al. 2023). The AtmoSOFAR channel is an atmospheric equivalent of the Sound Fixing and Ranging (SOFAR) channel in the ocean (Lawrence 1999), and it can span between 10 – 40 km altitude (Albert, Bowman et al. 2023).

Figure 7 shows the presence of the AtmoSOFAR channel between the source at 35 km altitude and I46RU. Further to this, in their study of elevated sonic boom sources, Nippress and Green (2017) noted that the acoustic energy trapped in the elevated duct could 'leak', thereby enabling the signal detection at ground-based stations. They proposed diffraction out of the waveguide as one of the possible mechanisms. The fact that the signal at I46RU was not theoretically predicted but was obviously detected provides compelling evidence that the upwind propagation and subsequent detection of acoustic energy was facilitated by the leaky AtmoSOFAR channel. This study demonstrates that the intricate interplay between infrasound wave propagation through the AtmoSOFAR channel (Albert, Bowman et al. 2023) and diffraction out of the waveguide (Nippress & Green 2017) could results in unexpected signal detections at distant stations.

Events like this one also attest to the importance of having accurate atmospheric specifications, and the need to continuously refine propagation models. Moreover, future observations of bolides from an elevated vantage point using stratospheric balloons hold the potential to significantly enhance detection efficiency compared to ground-based sensing methods (Young, Bowman et al. 2018; Bowman 2021; Albert, Bowman et al. 2023). Because balloons float in the region of the atmosphere where the AtmoSOFAR channel forms, these sensing platforms could potentially detect signals from high-altitude sources such as bolides or supersonic vehicles at very large distances. Considering that balloons are carried by winds, the noise levels have been found to be lower than that in the ground-based sensors (Krishnamoorthy et al. 2020), although some instances of sporadic noise of unknown origin has been noted (Silber et al. 2023b).

Pilger, Ceranna et al. (2015) examined the capability of the IMS network to detect energetic events such as the Chelyabinsk superbolide. They found that directivity can have a significant influence on signal detection efficiency, which correlated with a cylindrical line source geometry of the bolide. Strictly from the perspective of orientation, the Tajikistan bolide did not have a favorable geometry relative to the stations. However, compared to the Chelyabinsk event which was very shallow, the Tajikistan bolide entered at a steep angle.

Fragmentation events can complicate the usual cylindrical line source approximation. In addition to a cylindrically expanding blast wave, there is also a spherically expanding point source explosion wave; these need to be considered as different systems in terms of attenuation and overpressure decay. Depending on the bolide velocity, downward momentum can be significant, giving rise to a complicated mixture of the cylindrical and the spherical (or quasi-spherical) blast waves (e.g., ReVelle 1974). Here, observational evidence suggests that most of the energy was released during the main fragmentation event, likely at 35 km altitude. The steep rise in visible light emission seen in the light curve suggests that the initial gross fragmentation was highly energetic. In relative terms, the two smaller flares that followed are insignificant compared to the main fragmentation event. Their visible light intensity was an order of magnitude smaller. Therefore, considering that signal periods measured at the two stations are the same to within 2%, it is reasonable to conclude that a spherically expanding blast wave was the dominant mode of shock production.

The total infrasound-derived yield using Eq. (4a), based on the signal period averaged between the two stations, is 0.19 ± 0.01 kt. For comparison, the total impact energy published by CNEOS was estimated at 0.36 kt. The energy estimate using Eq. (5b) is 0.17 ± 0.01 kt. The relations from Gi and Brown (2017) give larger values by nearly a factor of three. Eqs. (6a) and (6b), result in yields of 0.50 ± 0.01 kt and 0.59 ± 0.01 kt, respectively. The apparent difference in energy estimate is likely due to the way the relations were calibrated. Eq. (4a) is based on nuclear explosion data, where the burst altitude is much lower than that of bolides. Eqs. (5a) and (5b), on the other hand, were calibrated using the USG-derived energy, which is based on the integrated light curve, and represents the energy at the top of the atmosphere. Bolides deposit energy at much higher altitudes than nuclear explosions, and therefore, all things being equal, one would expect that the bolide periods would be larger. Gi and Brown (2017) suggest that the altitude effect might be countered by the fact that the cylindrical line source type of energy deposition (per unit path length) will have smaller periods. Nevertheless, these energy estimates are still in good agreement. The infrasonic mass then is between 6.6 and 23.5 tons. Assuming a chondritic composition (3400 kg/m$^3$) (Consolmagno S.J & Britt 1998), the object radius was between 0.78 and 1.18 meters. This estimate agrees well with those published by Konovalova, Madiedo et al. (2013) and Peña-Asensio, Trigo-Rodríguez et al. (2022).

## 6. Conclusion

Infrasound is a remote sensing modality that can provide continuous global monitoring, which can be leveraged towards planetary defense and advancing non-proliferation efforts. Infrasound detections of large bolides play a pivotal role in that undertaking; however, detailed reports of these events are still few and far between in literature. This work describes infrasonic detection and detailed analysis of the Tajikistan bolide which impacted on 23 July 2008. The Tajikistan bolide was detected by two infrasound stations of the IMS network at distances 1530 and 2130 km. The dominant signal periods were consistent across the stations but the overall frequency content was quite different. Propagation paths to I46RU were not predicted by the model, yet, the signal was detected, albeit with a much lower frequency content. While atmospheric specifications given by various models are considered generally robust, they do not include fine-scale perturbations or other local effects that might occur. Here, the acoustic energy was trapped in a weak but leaky AtmoSOFAR duct. The intricate interplay between infrasound wave propagation through the AtmoSOFAR channel and diffraction out of the waveguide could facilitate upwind signal detections, as evidenced in this work. The infrasound signal analyses at the two stations, combined with the likely source altitude derived from propagation modeling at I31KZ, suggest that the shock originated at the point of the main breakup, in agreement with observational evidence. Furthermore, it has been determined that the mode of shock production was a spherical blast resulting from gross fragmentation. The updated energy estimate, based on the signal period, is between 0.17 and 0.51 kt, suggesting a mass of 6.6–23.5 tons. Assuming chondritic composition, the object radius was 0.78–1.18 m, consistent with previous estimates. Future observations of bolides from an elevated vantage point using stratospheric balloons could offer additional constraints that might not be achievable through ground-based infrasound sensing. This is especially important in the context of propagation channels that might trap the acoustic energy at altitude, preventing the signals from efficiently reaching the ground.

**Acknowledgements:** Sandia National Laboratories is a multi-mission laboratory managed and operated by National Technology and Engineering Solutions of Sandia, LLC (NTESS), a wholly owned subsidiary of Honeywell International Inc., for the U.S. Department of Energy's National Nuclear Security Administration (DOE/NNSA) under contract DE-NA0003525. This written work is authored by an employee of NTESS. The employee, not NTESS, owns the right, title, and interest in and to the written work and is responsible for its contents. Any subjective views or opinions that might be expressed in the written work do not necessarily represent the views of the U.S. Government. The publisher acknowledges that the U.S. Government retains a non-exclusive, paid-up, irrevocable, world-wide license to publish or reproduce the published form of this written work or allow others to do so, for U.S. Government purposes. The DOE will provide public access to results of federally sponsored research in accordance with the DOE Public Access Plan.

**Funding:** This work was supported by the Laboratory Directed Research and Development (LDRD) program at Sandia National Laboratories, a multimission laboratory managed and operated by National Technology and Engineering Solutions of Sandia, LLC., a wholly owned subsidiary of Honeywell International, Inc., for the U.S. Department of Energy's National Nuclear Security Administration under contract DE-NA-0003525.

**Conflict of interest:** The author declares no conflict of interest.

Silber, E. A. (2024) *Utility of infrasound towards global monitoring of extraterrestrial impacts: A case study of the 23 July 2008 Tajikistan bolide*, The Astronomical Journal, accepted manuscript, DOI: 10.3847/1538-3881/ad47c3
Sansom, E. K., et al. 2019, The Astrophysical Journal, 885, 115.

Silber, E. A., M. Boslough, W. K. Hocking, M. Gritsevich & R. W. Whitaker 2018, Advances in Space Research, 62, 489.

Silber, E. A., D. C. Bowman & S. Albert 2023a, Atmosphere, 14, 1473.

Silber, E. A., D. C. Bownman & M. TRonac Giannone 2023b, Remote Sensing, Accepted for publication.

Silber, E. A. & P. Brown 2019. in. ed. A. Le Pichon, E. Blanc and A. Hauchecorne. (Cham: Springer International Publishing), 939.

Silber, E. A., A. Le Pichon & P. G. Brown 2011, Geophys Res Lett, 38, L12201.

Silber, E. A., D. O. ReVelle, P. G. Brown & W. N. Edwards 2009, J Geophys Res, 114, E08006.

Southerland, L. C. & H. Bass, E. 2004, J Acoustic Soc Am, 115, 1012.

Swinbank, R. & A. A. O'Neill 1994, Monthly Weather Review, 122, 686.

Tagliaferri, E., R. Spalding, C. Jacobs, S. P. Worden & A. Erlich 1994, Hazards due to Comets and Asteroids, 24, 199.

Trigo-Rodríguez, J. M. 2022, Asteroid Impact Risk, Impact Hazard from Asteroids and Comets (Cham, Switzerland: Springer

Tsikulin, M. (1970). Shock waves during the movement of large meteorites in the atmosphere. Nat. Tech. Inform. Serv., Springfield, Va DTIC Document AD 715-537.

Webb, W. L. 1966, Structure of the stratosphere and mesosphere, Vol. 9 (Academic Press (New York)

Whipple, F. 1930, Quarterly Journal of the Royal Meteorological Society, 56, 287.

Young, C. J., E. P. Chael & B. J. Merchant Year, in, 24th Seismic Research Review (

Young, E. F., D. C. Bowman, J. M. Lees, V. Klein, S. J. Arrowsmith & C. Ballard 2018, Seismological Research Letters, 89, 1497.

Zinn, J., P. J. O'Dean & D. O. ReVelle 2004, Advances in Space Research, 33, 1466.
29

SAND2024-05683O